# Atmospheric Conditions, Electric Fields, and Thunderstorm Ground Enhancements at Aragats

**A.Chilingarian*, N.Manukyan, Q. Piliposyan**
**A.I.Alikhanyan National Lab (Yerevan Physics Institute)**
**Alikhanyan Brothers 2, Yerevan, Armenia, AM0036.**

*Corresponding author:
E-mail: chili@aragats.am

## Abstract

The Aragats High-Altitude Research Station (3200 m a.s.l.) combines meteorological, atmospheric-electric, and particle-flux measurements, providing a unique setting for investigating thunderstorm ground enhancements (TGEs) and recent environmental variability. We analyze one-minute observations from 2012–2025, first focusing on the meteorological and electric-field conditions controlling TGEs and then on trends in air temperature and daytime solar radiation. From an archive of 610 TGEs, a quality-controlled subset of 284 events with particle flux enhancements ≥10% was selected. Comparisons of May, August, and October show that strong near-surface electric fields alone are insufficient for frequent TGEs. May combines strong-field conditions with low cloud bases and shows a pronounced inverse relation between TGE occurrence and cloud-base distance. Thus, TGE occurrence is jointly controlled by thundercloud electrification and cloud-to-detector distance, while the recent warming tendency cannot be attributed to increased measured short-wave radiation. The annual mean air temperature shows a statistically significant positive trend, whereas daytime solar radiation shows a significant negative trend. Thus, warming at Aragats cannot be explained solely by variations in solar radiation.

## Plain Language Summary

Thunderstorms can accelerate electrons to high energies, producing bursts of electrons and gamma rays detected at ground level, known as thunderstorm ground enhancements (TGEs). We studied the weather and electric-field conditions accompanying these events at the Aragats Research Station, 3200 m above sea level. Of 284 selected TGEs, most occurred in spring, when thundercloud bases were often close to the station. Summer cloud bases were much higher, making it difficult for electrons and even some gamma rays to reach the detectors. A comparison of May, August, and October shows that a strong electric field measured at the ground is not sufficient by itself: frequent TGEs occur when a strong field and a low cloud base are present together. We also examined the 2012–2025 temperature and sunlight records. Air temperature increased by about 1.16 °C per decade, while measured daytime solar radiation decreased. Statistical tests show that both tendencies are robust. The Aragats measurements therefore link local atmospheric conditions with both high-energy thunderstorm processes and recent environmental change.

## Key Points

• Low cloud bases and strong near-surface electric fields act together to control TGE occurrence at Aragats.
• Spring and autumn TGEs occur under much lower cloud bases than summer events, favoring electron detection.
• From 2012–2025, air temperature increased by 1.16 °C per decade while daytime solar radiation significantly decreased.

## 1. Introduction

Mountain environments are among the most sensitive to environmental variability because meteorological conditions are strongly influenced by altitude, topography, snow cover, cloud formation, atmospheric circulation, and radiative processes. However, interpreting mountain records requires careful consideration of local terrain, exposure, and circulation conditions, because observed changes often reflect a combination of regional-scale forcing and site-specific processes.

The South Caucasus remains substantially underrepresented in the international mountain-climate literature compared with the Alps, Himalayas, or Andes. Although substantial environmental changes have been reported across Europe and Asia in recent decades, long-term high-altitude observations from the Caucasus remain scarce. As a result, the region's response to recent environmental variability is less well documented than that of the Alps, Himalayas, or Andes. Armenia lies at the intersection of several major circulation influences, including westerly flows from the Mediterranean, continental air masses from Eurasia, and occasional subtropical intrusions from the Middle East. This variability drives pronounced seasonal and interannual fluctuations in temperature, precipitation, and atmospheric transparency.

The Aragats Research Station, operated by the Yerevan Physics Institute (YerPhI), is located at 3200 m on the southern slope of Mount Aragats, the highest mountain in Armenia (4090 m a.s.l.). Mount Aragats consists of four volcanic summits surrounding a large crater and is a dominant topographic feature of the Armenian Highlands. The station experiences prolonged snow cover, up to 2 m thick, strong solar irradiance, frequent cloud formation, rapid weather changes, and pronounced seasonal contrasts.  The location and long-standing observational infrastructure of the Aragats station are documented by the Cosmic Ray Division (2026a) and Chilingarian et al. (2024c).
Figure 1 provides the geographical and observational context of the present study. Panel A shows Armenia and the network of DAVIS meteorological stations operated by YerPhI, including Aragats, Nor Amberd, Yerevan, Sevan, and Dilijan. Because identical instrumentation has been used at these sites, the network provides an opportunity for future regional comparisons and independent validation of observed environmental tendencies.

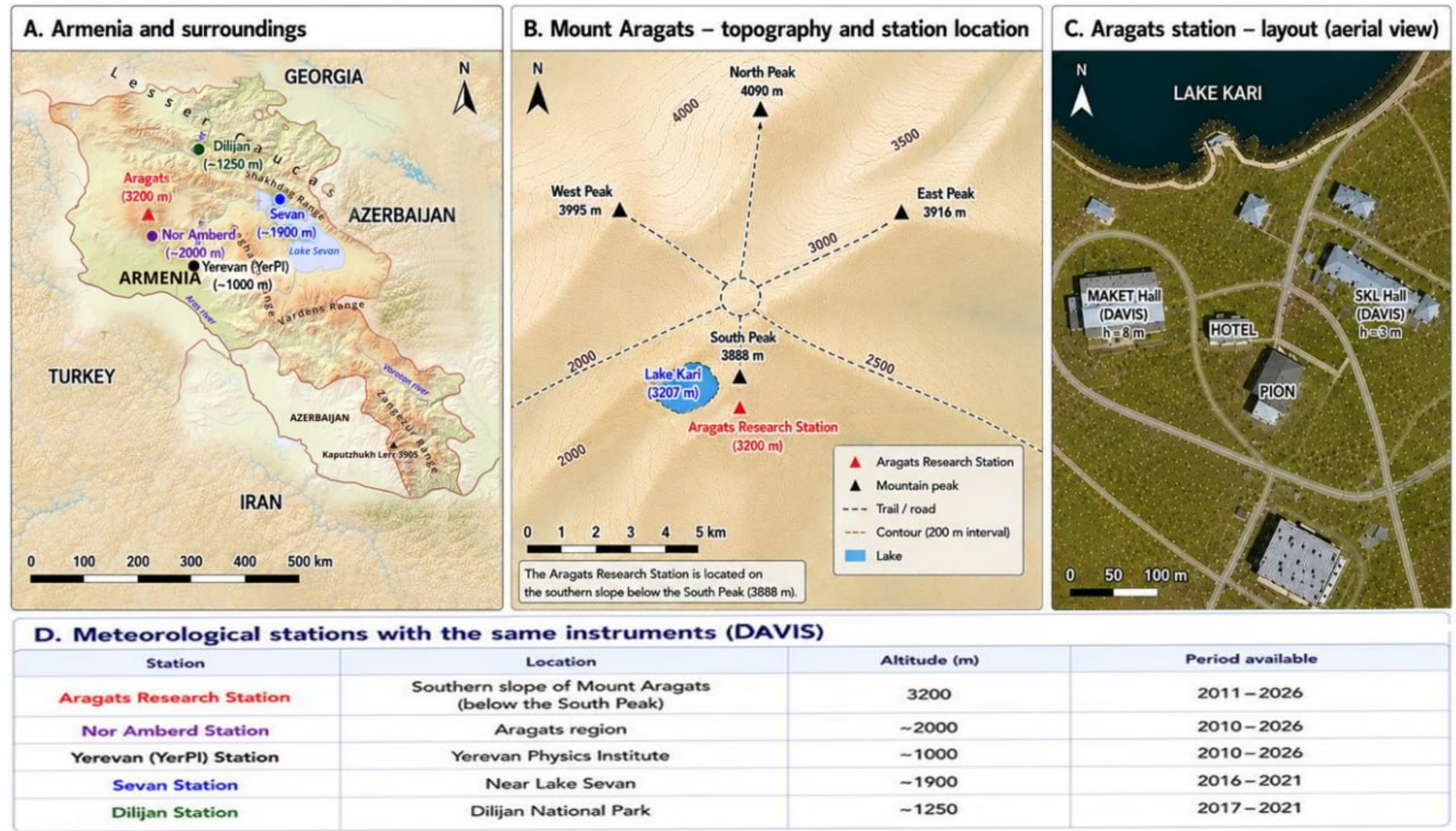

| Station | Location | Altitude (m) | Period available |
|---|---|---|---|
| Aragats Research Station | Southern slope of Mount Aragats (below the South Peak) | 3200 | 2011 – 2026 |
| Nor Amberd Station | Aragats region | ~2000 | 2010 – 2026 |
| Yerevan (YerPI) Station | Yerevan Physics Institute | ~1000 | 2010 – 2026 |
| Sevan Station | Near Lake Sevan | ~1900 | 2016 – 2021 |
| Dilijan Station | Dilijan National Park | ~1250 | 2017 – 2021 |

**Figure 1. Geographical location, topographic setting, and observational infrastructure of the Aragats Research Station.**

Panel A shows Armenia and the network of DAVIS meteorological stations operated by CRD/YerPhI. Panel B presents the topography of Mount Aragats, including its four peaks, Lake Kari, and the position of the Aragats Research Station. Panel C shows the station layout and locations of the DAVIS meteorological stations at the MAKET and SKL buildings. Panel D summarizes the operation periods of the DAVIS stations used for regional comparison.

The Aragats station is part of the Armenian Geophysical Network, operated by the Cosmic Ray Division of the Yerevan Physics Institute. During the period covered in this study, identical DAVIS meteorological stations operated at Aragats, Nor Amberd, Yerevan (YerPhI), Sevan, and Dilijan. The Sevan and Dilijan stations ceased operations in 2021, whereas Aragats, Nor Amberd, and Yerevan continue to provide observations. The availability of measurements from multiple sites with identical instrumentation offers a valuable opportunity for future investigations of regional-scale environmental variability and for independent validation of trends observed at Aragats. The observatory is also a member of the Virtual Alpine Observatory (VAO) initiative, which promotes long-term environmental monitoring and data sharing among mountain observatories. This affiliation places Aragats within an international framework of high-altitude observational sites and facilitates future comparisons of measurements from the South Caucasus with those from other mountain regions. Continuous meteorological measurements have been available at Aragats since 2011. The archive contains one-minute observations of air temperature, humidity, pressure, wind, precipitation, solar radiation, and ultraviolet radiation. The present analysis uses 14 calendar years, 2012–2025; nevertheless, the multivariate observations permit characterization of temperature and radiation trends and their relationships with high-energy thunderstorm phenomena.

Preliminary inspection of the Aragats archive indicates that near-surface air temperature shows a positive trend over the analyzed period, whereas measured daytime solar radiation does not increase and may even decrease. If confirmed, this behavior suggests that temperature variability at Aragats cannot be explained solely by changes in incoming short-wave radiation. The present study, therefore, serves as a first step toward a broader investigation of environmental variability in the Armenian Highlands. The DAVIS Vantage Pro2 Plus instrumentation and its operation at Aragats are documented by the Cosmic Ray Division (2026b) and Davis Instruments (2021).

The Aragats facilities are widely used in solar physics and solar-terrestrial relations (Chilingarian et al., 2026; Sargsyan and Chilingarian, 2026). This opens perspectives for studying the Sun-Earth system as a complex system, determined by the variability of solar output energy and the energy transfer from the Sun to the Earth.
The experimental program on Aragats included research on electron and positron accelerators in thunderous atmospheric conditions and on the structure of the atmospheric electric field (AEF). These aspects of atmospheric science are related to overall meteorological research, and both are compiled in the same Aragats multiyear database.

## 2. Thunderstorm ground enhancements and meteorological conditions at Aragats

A thunderstorm ground enhancement (TGE) is a transient increase in the fluxes of electrons and gamma rays detected at ground level during thunderstorms. TGEs originate mainly from relativistic runaway electron avalanches (RREAs, Gurevich et al., 1992) developing in strong atmospheric electric fields (AEFs) inside thunderclouds. The accelerated electrons produce bremsstrahlung gamma rays, which in turn produce electron-positron pairs, thereby developing an avalanche. When the cloud base and the acceleration region are sufficiently close to Earth's surface, energetic particles can reach the detectors. Because electrons are attenuated in the atmosphere much faster than gamma rays, special meteorological conditions are required to support electron registration.
Although C.T.R. Wilson predicted electron accelerators in thunderclouds at the beginning of the last century (Wilson, 1925), the negative results of measurements in South Africa (Schonland and Viljoen, 1933) set this field of research back for many decades and stand in stark contrast to present-day evidence for electron runaway, most notably at Mt. Aragats in Armenia (e.g., Chilingarian et al., 2010, 2011). The shortcomings in the early measurements and the rapid progress in this topic at the beginning of the 21st century were analyzed in Chilingarian et al. (2025), which highlighted the crucial influence of meteorological conditions on TGE detection, including the AEF and electric field geometry.

The near-surface electric field (NSEF), the local atmospheric electric field measured at the station by the network of Boltek EFM-100 field mills, can serve as a proxy for AEF, indicating thundercloud electrification, charge evolution, and nearby lightning activity. The EFM-100 instrument and the Aragats field-mill network are described by Boltek (2023) and Chilingarian et al. (2024c).
Particle fluxes are monitored by multiple detectors and spectrometers, while simultaneous temperature, dew-point temperature, and relative humidity are measured by DAVIS weather stations. For the meteorological analysis below, the total TGE archive of 610 events was restricted to 284 events recorded during 2012–2025 with a peak TGE enhancement of at least 10% (measured by the 3-cm-thick, 1 $m^2$-area plastic scintillator of the STAND3 detector, Chilingarian and Hovsepyan, 2023), relative humidity of at least 80%, and valid temperature and dew-point measurements. The particle detectors, their response channels, and the multivariate measurement infrastructure are described in Chilingarian et al. (2010, 2022, 2024c).

Panels (a)–(d) of Figure 2 connect the seasonal occurrence of TGEs with the local thermodynamic conditions that determine the distance between the cloud base and the particle detectors. The same quality-controlled sample is used in all panels, avoiding a comparison of partly different event sets. The event-strength threshold removes marginal particle enhancements, whereas the humidity requirement removes unrealistically large cloud-base estimates obtained when dry near-surface air is not representative of the thundercloud. Thus, the meteorological measurements quantify the atmospheric geometry controlling whether particles accelerated in the thundercloud can reach ground level.
The cloud-base distance was estimated as the surface-based lifting condensation level (LCL) above the 3200-m station using $H = 122(T - Td)$, where T and Td are the simultaneous near-surface air temperature and dew-point temperature in °C, and $T - Td$ is the temperature–dew-point spread (Lawrence, 2005; Samanta et al., 2020). This linear approximation follows from the faster cooling of a rising unsaturated air parcel than of its dew point and estimates the height at which the parcel first becomes saturated. It is therefore a proxy for the base of a nearby convective cloud, not a direct ceilometer measurement and not the cloud altitude above sea level.

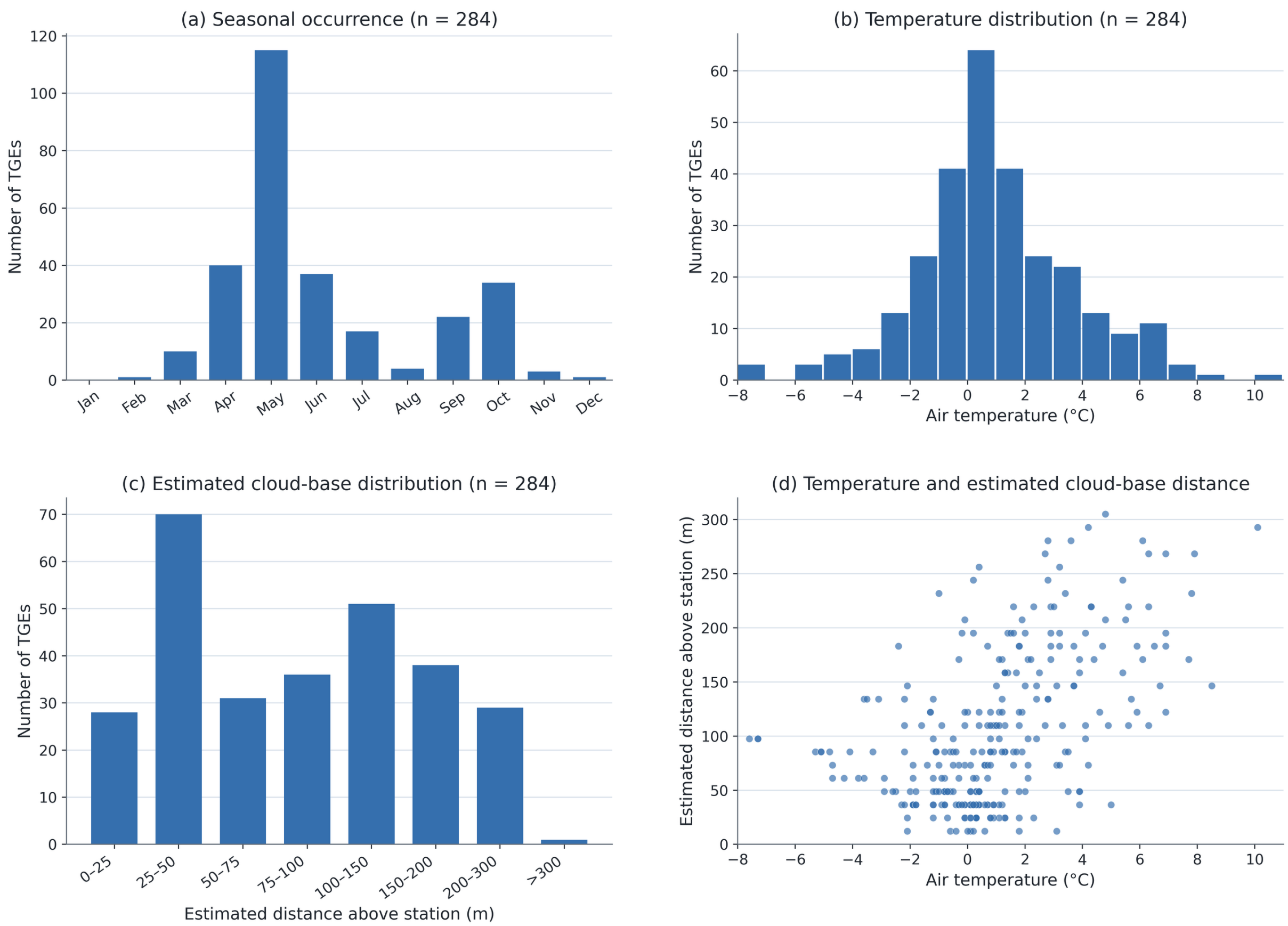

**Figure 2. Meteorological conditions for 284 TGEs recorded at the Aragats high-altitude station during 2012–2025. The sample is restricted to events with a peak STAND3 enhancement of at least 10%, relative humidity of at least 80%, and valid temperature and dew-point measurements. (a) Monthly distribution of the selected events. (b) Distribution of near-surface air temperature. (c) Distribution of the estimated cloud-base distance above the station, calculated as $H = 122(T - T_d)$. (d) Joint distribution of temperature and estimated cloud-base distance. The STAND3 threshold excludes marginal enhancements and is used only as an event-strength selection criterion, not as an electron-TGE classification criterion. The humidity requirement limits the cloud-base analysis to moist conditions in which the surface temperature and dew-point estimate is representative of the nearby cloud environment.**

The occurrence of TGEs at Aragats is strongly seasonal because it depends on the simultaneous presence of an electrified thunderstorm and a sufficiently short distance between the accelerating region and the detectors. Of the 284 selected TGEs, 165 (58.1%) occurred in meteorological spring, 58 (20.4%) in summer, 59 (20.8%) in autumn, and only 2 (0.7%) in winter (Panel a). The maximum occurred in May, with 115 events, followed by April with 40, June with 37, and October with 34 events.

Temperature, season, and estimated cloud-base distance are closely related (Panels b - d). In spring, the median temperature of TGE events was 0.1°C, and the median estimated cloud-base distance was 73.2 m. Autumn showed similar conditions, with median values of 1.1°C and 73.2 m, respectively. By contrast, summer TGEs occurred at a median temperature of 4.2°C and a median cloud-base distance of 158.6 m. An estimated cloud base below 100 m was found for 67.9% of spring TGEs and 66.1% of autumn TGEs, but for only 20.7% of summer events. Warmer conditions are accompanied by a greater distance between the cloud base and the station. Because the cloud-base estimate is calculated from the temperature–dew-point difference, this correlation should be interpreted as a descriptive meteorological relationship.

The meteorological differences between the seasons also help explain the particle composition of TGEs. Gamma rays are considerably more penetrating than electrons and can reach the detectors from comparatively greater distances (Chilingarian et al., 2010; Williams et al., 2023). Therefore, TGEs occurring under warm conditions and with a relatively high cloud base are expected to be predominantly gamma-ray events. Electrons suffer much stronger ionization losses and scattering in air; consequently, electron-rich TGEs require the lower boundary of the accelerating region to approach the station, typically to within several tens of meters (Chilingarian et al., 2017, 2022; Chilingarian et al., 2024a; 2024b). A cloud-base distance below approximately 50 m is therefore favorable for electron detection, although this value is a physically motivated descriptive threshold rather than a universal cutoff and is not sufficient by itself: the strength, polarity, extent, and duration of the atmospheric electric field are also essential.

The monthly results illustrate this distinction. In May, the median cloud-base distance was 73.2 m, and 46.1% of the selected TGEs occurred with an estimated cloud base below 50 m. In October, the median distance decreased to 54.9 m, and 50.0% of events occurred below 50 m. These conditions are particularly favorable for electron-rich TGEs. In June, however, the median cloud-base distance increased to 158.6 m, and only 5.4% of events occurred below 50 m. Thus, electron-rich June TGEs should be associated with exceptional episodes when the cloud base temporarily descends close to the station, whereas most summer TGEs should be gamma-dominated. In July and August, warmer air and higher cloud bases make the survival of electrons to detector level especially unlikely.

These results provide the physical link between the meteorological and particle measurements. Seasonal temperature and humidity control thundercloud development and the cloud-to-detector geometry. This geometry, in turn, determines the atmospheric path traversed by accelerated particles and therefore the relative probability of detecting electrons and gamma rays at ground level. The TGE amplitude used to select the present sample measures event strength and does not classify the electron-to-gamma ray ratio; identification of electron-rich TGEs requires independent charged- and neutral-particle measurements, which were made at Aragats with SEVAN and ASNT detectors (Chilingarian et al., 2022).

## 3. Cloud-base distance, strong near-surface electric fields, and TGE occurrence

In this section, we examine how cloud-base distance varies during periods of strong atmospheric electric fields during years of observation and how this relationship affects TGE occurrence. The analysis was restricted to May, August, and October during 2012–2024, as these months capture the main characteristics of thunderstorm activity at Aragats, when electron accelerators emerge in thunderclouds. Strong electric-field conditions were selected from the EFM-100 field mill record, located on the roof of the MAKET experimental hall (see Fig. 1), using the criterion |NSEF| > 5 kV m⁻¹. The NSEF strength is used to identify electrically active periods and to follow storm-charge evolution. Cloud-base distance during electrically active periods is a key factor in registering TGEs. Strong electric fields provide the acceleration mechanism, whereas low cloud bases minimize particle flux attenuation between the acceleration region and the detectors. The largest TGEs occur when both conditions are satisfied simultaneously. In August, the large cloud-base distance strongly suppresses the observation of electron-rich TGEs at the station level. In contrast, in May and October, when clouds are very near the surface, most electron TGEs are registered (Chilingarian et al., 2024b).

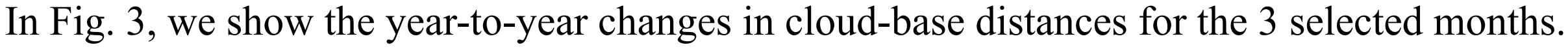

In Fig. 3, we show the year-to-year changes in cloud-base distances for the 3 selected months.

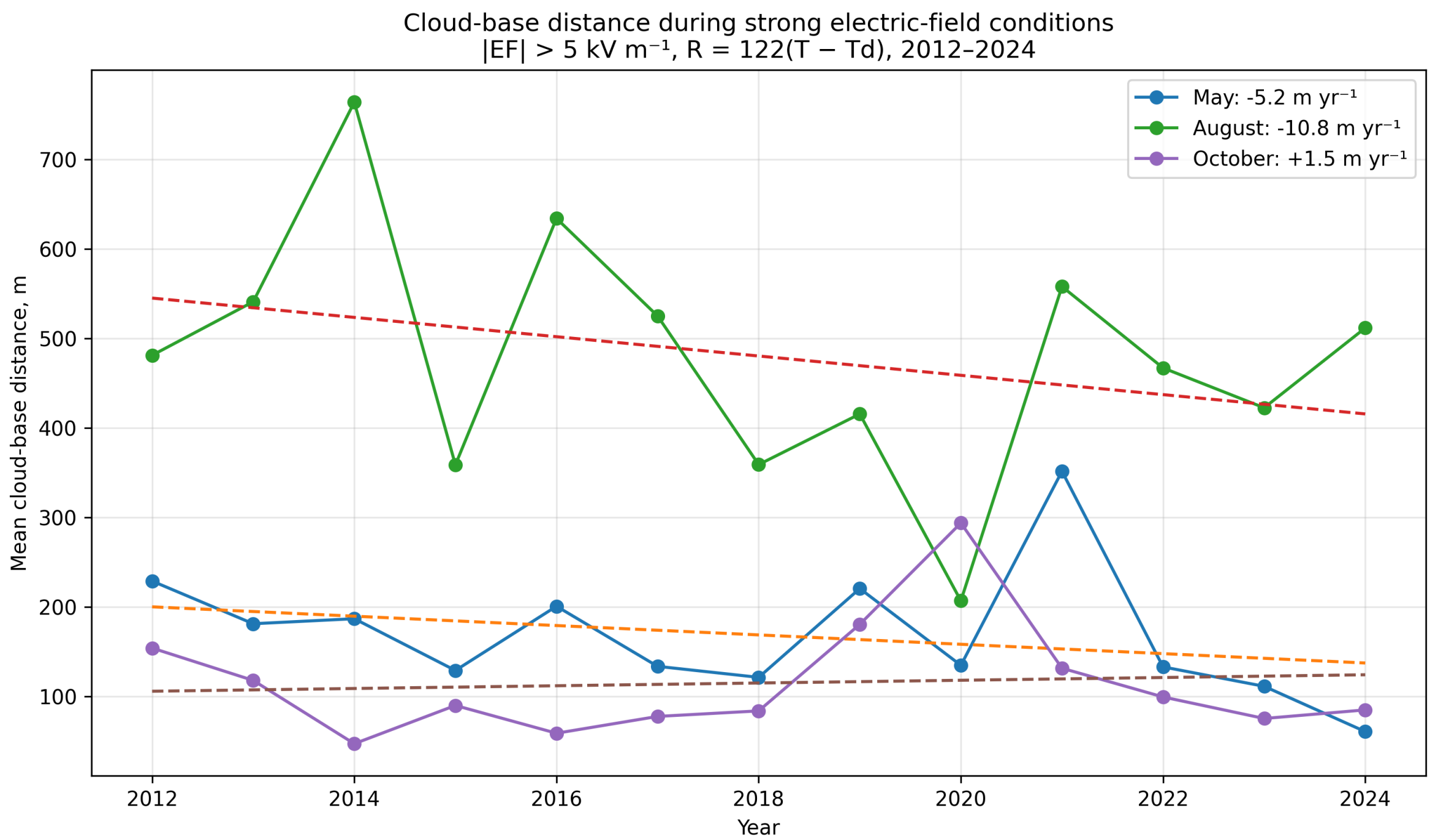

**Figure 3. Annual mean cloud-base distance during strong electric-field conditions (|NSEF| > 5 kV m⁻¹) for May, August, and October during 2012–2024. Dashed lines show linear fits. The figure compares the month-to-month behavior of electrically active periods based on the temperature–dew-point spread.**

Figure 3 highlights the key meteorological differences among the three months. The mean cloud-base distance during strong-field conditions is approximately 480 m in August, compared with 168 m in May and 115 m in October. The figure also indicates that interannual variability is large relative to long-term trends. Therefore, changes in TGE frequency cannot be attributed simply to a monotonic lowering of cloud bases. Instead, TGE occurrence depends on how often years feature low cloud bases coinciding with strong atmospheric electric fields.

A notable result is the persistent separation between August and the other two months. This separation offers a natural explanation for the scarcity of strong TGEs in summer, despite thunderstorms and enhanced electric fields.

Figure 4 compares cloud-base distances from all meteorological observations with those during periods when |EF| > 5 kV $m^{-1}$. The conditional curve is consistently lower than the curve derived from the full dataset. The persistent separation between the two curves indicates that strong electric fields preferentially develop as cloud systems approach the station.

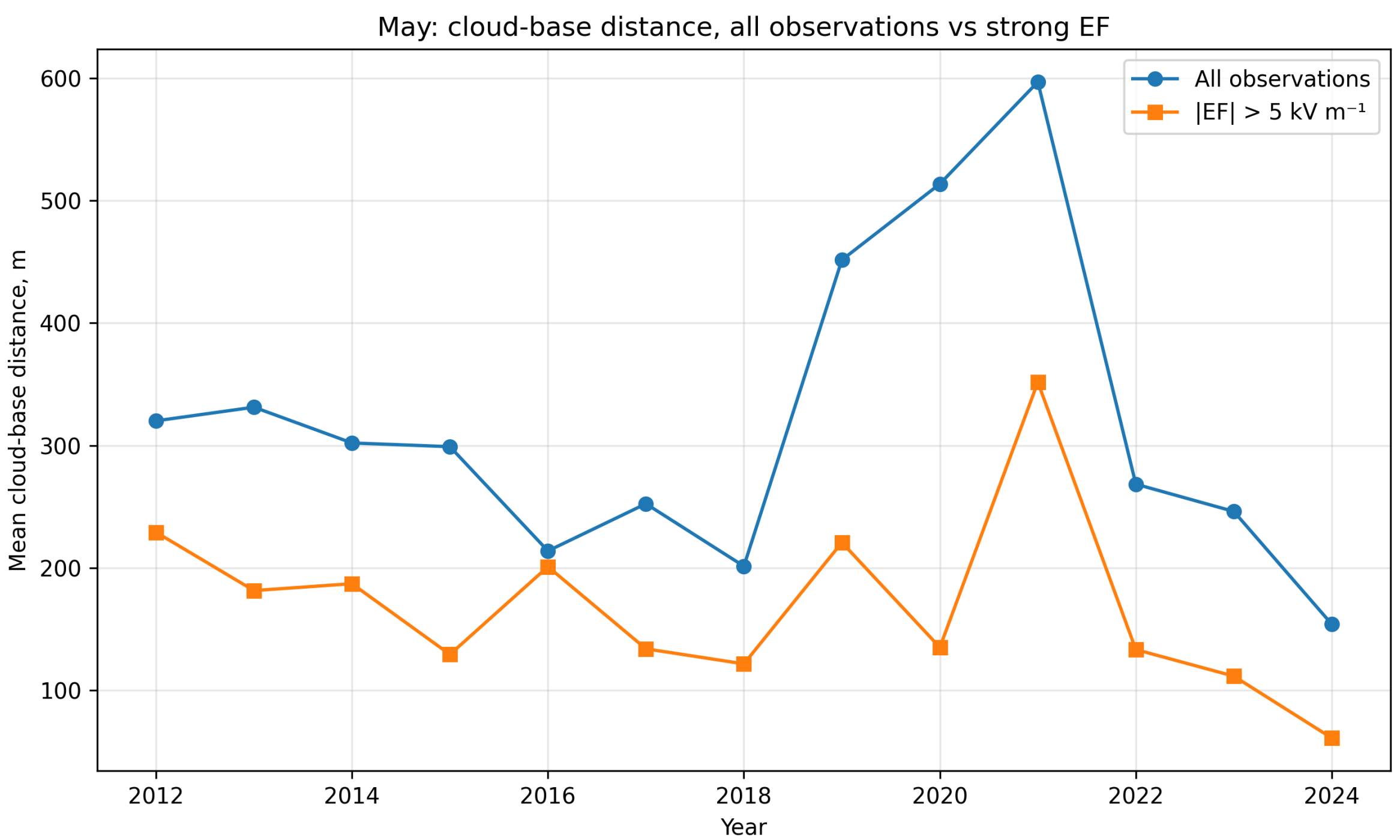

**Figure 4. May: comparison of annual mean cloud-base distance calculated from all observations and from the subset satisfying |NSEF| > 5 kV $m^{-1}$. The persistent separation between the two curves indicates that strong electric fields tend to occur when cloud systems approach the station.**

This coupling increases the likelihood that avalanche electrons and bremsstrahlung gamma rays generated within the cloud can reach the detectors.

Figure 5 summarizes the relationship between cloud-base distance and TGE occurrence over the three months. The strongest relationship occurs in May (panel a), with a correlation coefficient of $r = -0.668$. Years with lower cloud bases tend to have more TGEs. The years with the largest numbers of strong TGEs (2018, 2023) all correspond to cloud-base distances below 120 m. The ≥10% threshold defines the quality-controlled 284-event sample used in Figure 2, whereas the >20% threshold is used in Figure 5 to count comparatively strong TGEs.August (panel b) presents a fundamentally different situation. Although strong electric-field intervals occur, the cloud base remains far above the station, with mean distances around 480 m. Only one TGE with an enhancement exceeding 20% was recorded during the entire observation period. August, therefore, serves as a natural control sample, demonstrating that strong atmospheric electric fields alone are insufficient to produce frequent TGEs when cloud bases remain high.

In October (panel c), the correlation coefficient remains negative (r = −0.316), indicating the same physical tendency observed in May; however, the relationship is weaker because October contains fewer thunderstorms and substantially fewer strong-field intervals than May.

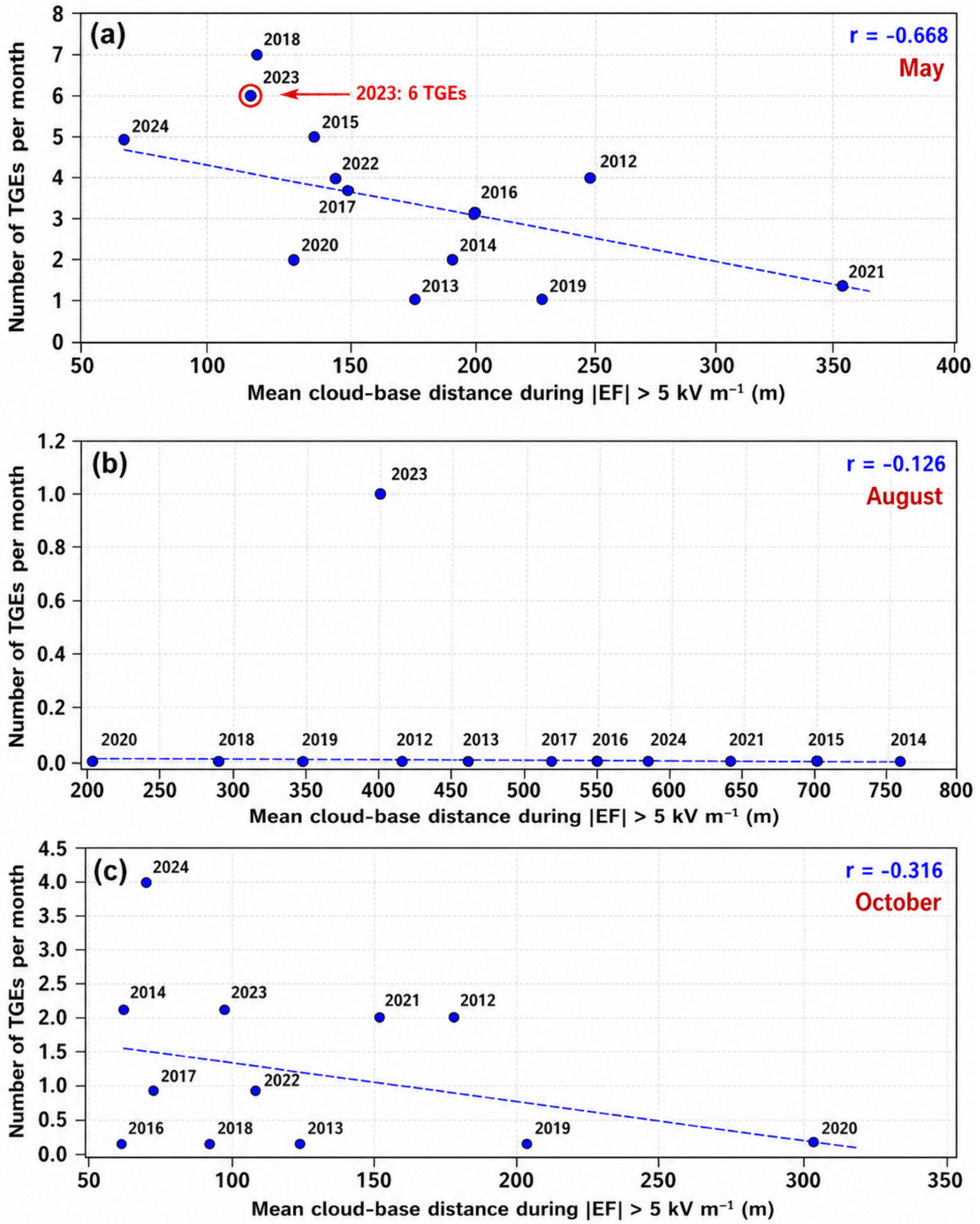

**Figure 5. Annual number of TGEs with STAND3 enhancements >20% versus mean cloud-base distance during periods with |EF| > 5 kV m⁻¹. Panels (a), (b), and (c) correspond to May, August, and October, respectively. Pearson correlation coefficients are shown in each panel.**

Taken together, the results indicate that neither cloud base distance nor the atmospheric electric field alone is sufficient to explain the occurrence of TGEs. Strong electric fields provide the acceleration mechanism, whereas low cloud bases minimize attenuation between the acceleration region and the detectors. The highest TGE frequencies occur when both conditions are met simultaneously. August provides numerous strong-field intervals but very few TGEs because cloud bases remain high. May combines frequent strong-field conditions with relatively low cloud bases and consequently exhibits the largest number of TGEs. This behavior supports the interpretation that TGE occurrence at Aragats is controlled by the combined action of cloud geometry and atmospheric electric fields rather than by either parameter individually.

## 4. Air temperature and daytime solar irradiance at Aragats, 2012–2025

In this section, we investigate trends in the meteorological time series (outdoor temperature and solar radiation) measured at Aragats during 2012-2025. Trend estimates are based on ordinary least-squares linear (OLS) regression. OLS slope denotes the slope of a straight line fitted to the annual means as a function of year. Linear-trend estimation from annual meteorological aggregates follows standard climatological statistical practice (Wilks, 2019).
Data quality control applied here was as follows. Temperatures outside -50 to 50 °C were treated as nonphysical sensor values and excluded. Solar radiation was restricted to 0-2000 W/m$^2$. Daytime analyses were defined as periods with solar radiation above 20 W/m$^2$. Table 1 documents the amount of available data for each full year from 2012 to 2025.

**Table 1. Annual data completeness for the 2012-2025 years.**

| Year | Observed rows | Row completeness % |
|---|---|---|
| 2012 | 478437 | 90.778 |
| 2013 | 492025 | 93.612 |
| 2014 | 493679 | 93.927 |
| 2015 | 525247 | 99.933 |
| 2016 | 516620 | 98.023 |
| 2017 | 525185 | 99.921 |
| 2018 | 524416 | 99.775 |
| 2019 | 474011 | 90.185 |
| 2020 | 510557 | 96.873 |
| 2021 | 511980 | 97.409 |
| 2022 | 494745 | 94.130 |
| 2023 | 517772 | 98.511 |
| 2024 | 498836 | 94.649 |
| 2025 | 522279 | 99.368 |

Table 2 shows the annual mean temperature and mean daytime solar radiation for each complete year.

Row completeness in Table 1 is generally high: most years exceed 94%, with several near 100%. The lowest row completeness occurs in 2019, but even then, about 90% of expected rows are retained. The table shows that annual means are based on hundreds of thousands of 1-minute observations, not sparse or occasional measurements.

**Table 2. Annual mean temperature and mean daytime solar radiation for complete years 2012–2025.**

| Year | Mean air temperature (°C) | Mean daytime solar irradiance (W $m^{-2}$) |
|---|---|---|
| 2012 | -0.709 | 381.968 |
| 2013 | -1.383 | 383.175 |
| 2014 | -0.162 | 377.837 |
| 2015 | -0.190 | 380.762 |
| 2016 | -1.132 | 370.624 |
| 2017 | -0.471 | 389.526 |
| 2018 | 0.404 | 357.367 |
| 2019 | 0.306 | 387.964 |
| 2020 | 0.013 | 370.675 |
| 2021 | 0.861 | 347.512 |
| 2022 | 0.231 | 359.776 |
| 2023 | 0.295 | 364.026 |
| 2024 | 0.192 | 335.098 |
| 2025 | 0.618 | 362.620 |

The annual mean temperature varies from −1.38 °C in 2013 to +0.62 °C in 2025, demonstrating substantial interannual variability. The annual mean daytime solar radiation ranges from approximately 335 to 390 W $m^{-2}$. The trend analysis is performed using these annual aggregates rather than individual minute measurements. This approach prevents artificial inflation of statistical significance caused by the very large number of high-frequency observations. The average temperature for the years 2012-2025 shows a clear upward trend of 0.116 ± 0.029 °C per year ($p = 0.0019$), which equals 1.16 °C per decade, Fig. 6. Broadband daytime solar radiation shows a negative trend of -2.62 ± 0.79 W/$m^2$ per year ($p = 0.0061$) over 2012–2025, Fig. 7. This suggests that the increase in temperature is not simply due to stronger measured incoming solar radiation. In practical terms, the annual temperature trend is statistically robust; the decrease in solar radiation is statistically significant as well.

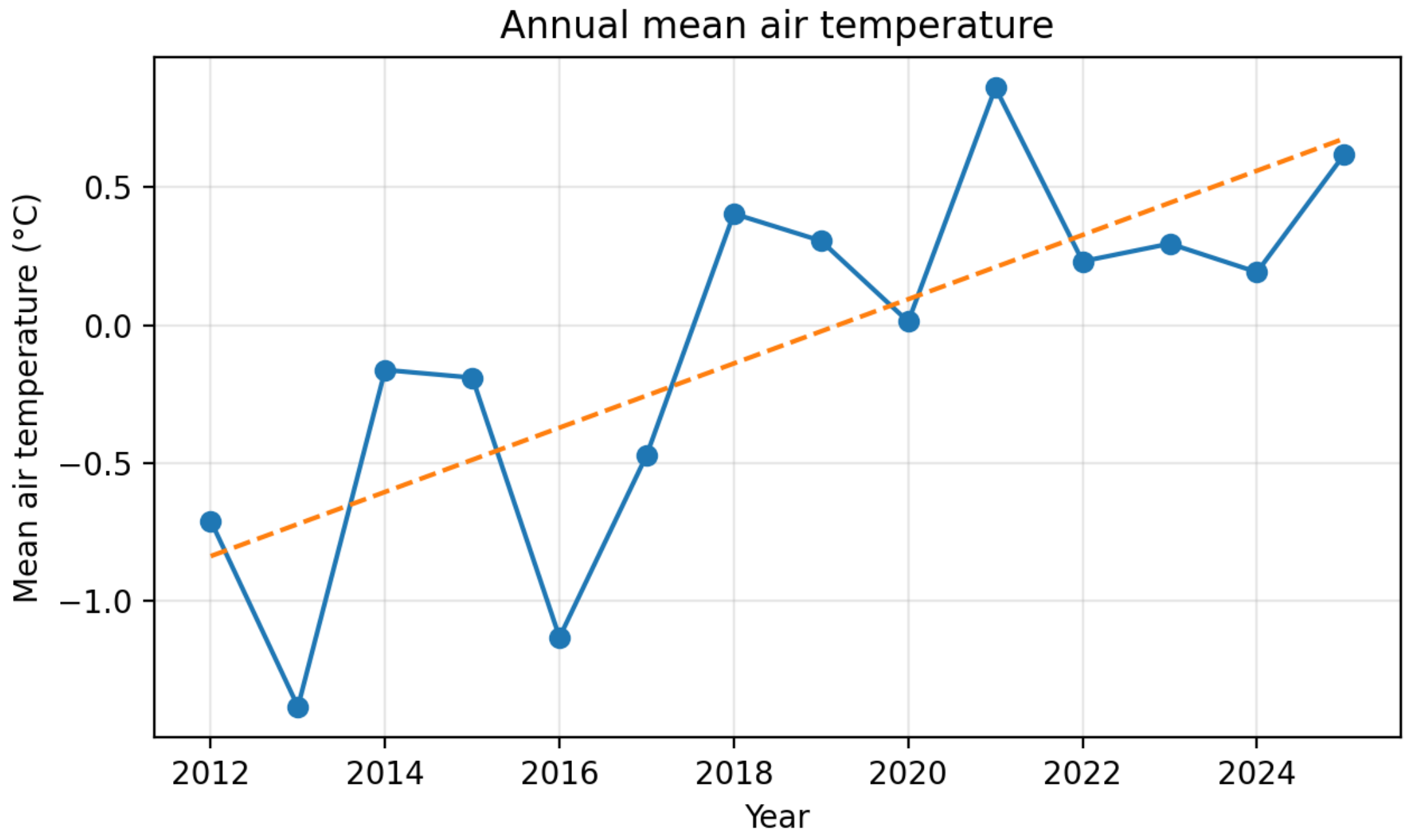

**Figure 6. Annual mean outside temperature. The dashed line is the linear fit over the full years 2012-2025. Despite large interannual variability, a positive tendency is present.**

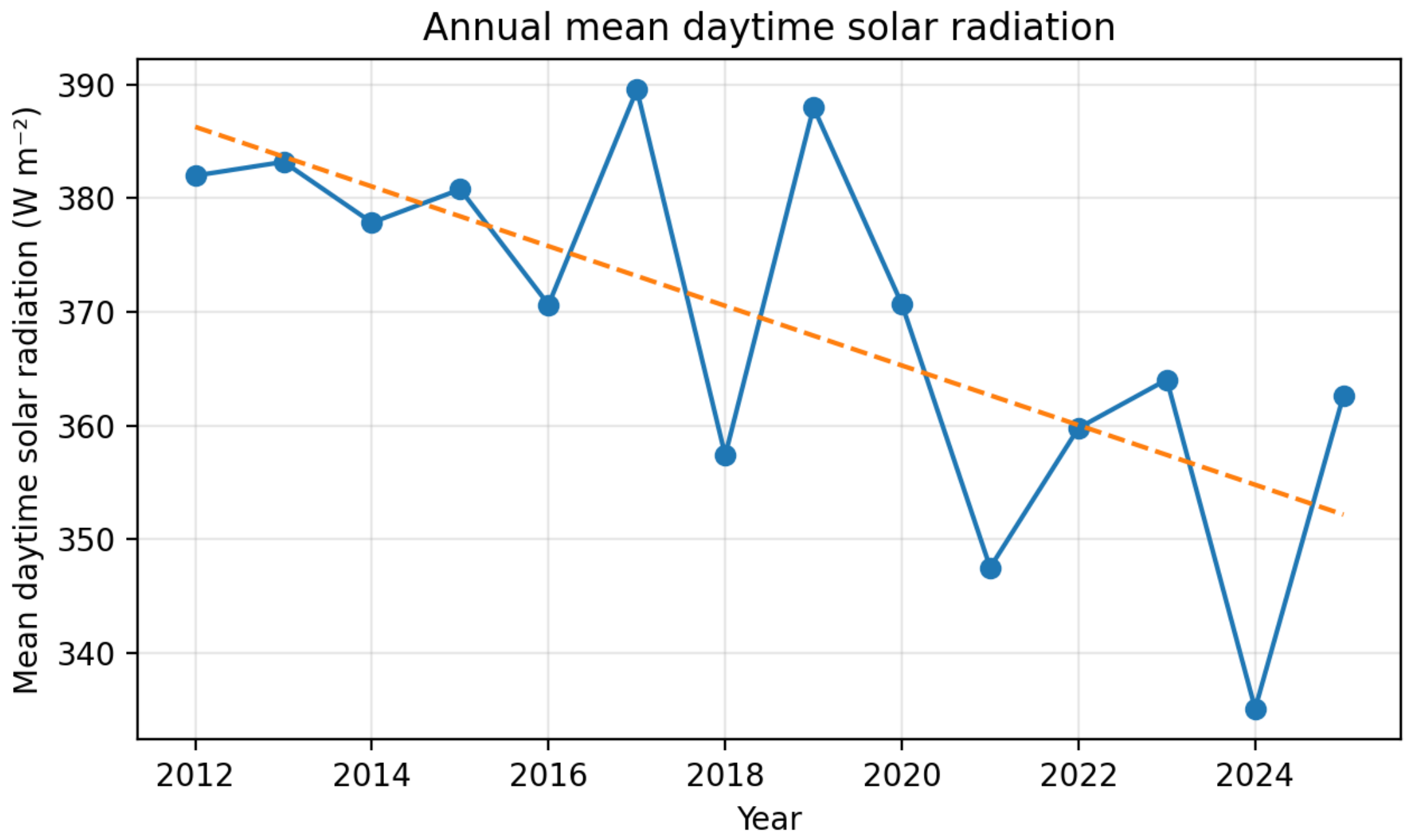

**Figure 7. Annual mean solar radiation. The dashed line is the linear fit over the full years 2012-2025. Despite large interannual variability, a negative tendency is present.**

To confirm annual trends, the sequence of fourteen annual values was resampled with replacement 20,000 times. For each of 20,000 bootstrap replicates, the 14 paired observations (year and annual value) were sampled with replacement, and an OLS slope was fitted to the resampled pairs. The median slope and the 2.5th and 97.5th percentiles were used as the bootstrap estimate and 95% confidence interval, respectively.  This resampling procedure follows the nonparametric bootstrap framework of Efron and Tibshirani (1993).
This procedure directly tests the stability of the annual trend against the limited length of the annual series. If the confidence interval remains entirely positive or entirely negative, the corresponding tendency is considered robust at the bootstrap level. If the interval overlaps zero, the sign of the trend is not stable enough for strong interpretation.
In Table 3, we present results from bootstrap tests. The OLS slope is obtained via ordinary least squares applied to the annual means. The bootstrap median is the median of the slope across 20,000 bootstrap resamples. $CI_{2.5}$ and $CI_{97.5}$ are the 2.5th and 97.5th percentiles of the bootstrap slope distribution, defining the 95% bootstrap confidence interval.

Table 3. **Bootstrap estimates of linear trends in annual mean air temperature and annual mean daytime solar radiation.**

| Variable | Trend (OLS slope) | Trend averaged over 20000 bootstrap resamples | $CI_{2.5}$ | $CI_{97.5}$ |
|---|---|---|---|---|
| Annual mean air temperature | 0.116 | 0.117 | 0.068 | 0.176 |
| Annual mean daytime solar radiation | -2.622 | -2.598 | -4.153 | -1.497 |
| Annual air temperature variance | 0.098 | 0.101 | -1.223 | 1.426 |
| Annual daytime solar-radiation variance | -458.333 | -455.660 | -767.144 | -145.416 |

Table 3 summarizes the central numerical test. The annual mean air temperature slope is +0.116 °C $yr^{-1}$, and the bootstrap median is nearly identical (+0.117 °C $yr^{-1}$). The 95% bootstrap interval is positive throughout, ranging from +0.068 to +0.176 °C $yr^{-1}$. The annual daytime solar-radiation slope is negative (-2.62 W $m^{-2}$ $yr^{-1}$), and the bootstrap interval remains fully negative, from -4.15 to -1.50 W $m^{-2}$ $yr^{-1}$.

The bootstrap analysis shows that the annual air temperature trend remains positive under bootstrap resampling, and its 95% confidence interval does not include zero. It justifies retaining the statement that the Aragats air temperature record shows a recent positive air temperature tendency over the complete years 2012–2025.
Bootstrap tests also confirm a decrease in daytime global solar radiation. Thus, the air temperature increase is not accompanied by stronger measured daytime short-wave radiation.

The result points to more complex mountain-atmosphere processes, including cloudiness, aerosol or dust effects, snow cover and albedo, boundary-layer structure, and regional circulation.

The test results in Table 4 assess whether the main trends are driven by rare extreme measurements.

Table 4. **Sensitivity of annual slopes to removal of extreme observations.**

| Sample | Variable | Slope per year |
|---|---|---|
| Full data | Air temperature | 0.116 |
| Full data | Daytime solar radiation | -2.622 |
| 0.5% tails removed | Air temperature | 0.108 |
| 0.5% tails removed | Daytime solar radiation | -2.382 |
| 1% tails removed | Air temperature | 0.101 |
| 1% tails removed | Daytime solar radiation | -2.367 |

The full air temperature slope is +0.116 °C $yr^{-1}$. After removing the most extreme 0.5% from both tails of the distribution, the slope remains positive (+0.108 °C $yr^{-1}$). After removing 1% from both tails, it remains positive (+0.101 °C $yr^{-1}$). The magnitude is slightly reduced, as expected when the coldest and warmest minutes are removed, but the sign and main conclusion are unchanged. The daytime solar-radiation slope also remains negative after trimming. Therefore, the principal results are not artifacts of isolated outliers, gross spikes, or rare extreme minutes. Thus, the bootstrap and trimming tests provide a strong data-quality argument. The annual air temperature increase and the daytime solar-radiation decrease are robust to resampling and outlier removal, confirming a recent warming tendency in the 2012–2025 Aragats record. The combined result demonstrates that the observed positive temperature tendency occurred despite a simultaneous decrease in measured daytime solar radiation.

## 5. Discussion and Conclusions

The combined particle, electric-field, and meteorological measurements at Aragats demonstrate that the observed occurrence of thunderstorm ground enhancements is governed by both thundercloud electrification and cloud-to-detector distance. Strong electric fields provide the conditions required for relativistic electron acceleration, whereas a low cloud base reduces the atmospheric attenuation of electrons and gamma rays before they reach the detectors. Neither condition alone adequately explains the observed TGE frequency.

The quality-controlled sample of 284 TGEs shows a pronounced seasonal dependence. Of these events, 58.1% occurred in spring, 20.4% in summer, 20.8% in autumn, and only 0.7% in winter. Spring and autumn TGEs occurred under similar conditions, with median estimated cloud-base distances of 73.2 m. In summer, the median distance increased to 158.6 m. Cloud bases below 100 m occurred during 67.9% of spring and 66.1% of autumn TGEs, but during only 20.7% of summer events. These differences explain why the conditions for detecting electrons are substantially more favorable in spring and autumn than in summer.

The May–August–October comparison provides independent support for this interpretation. During periods with $|\,\mathrm{NSEF}\,| > 5\ \mathrm{kV\,m^{-1}}$, the mean estimated cloud-base distance was approximately 168 m in May, 480 m in August, and 115 m in October. In May, the annual number of TGEs with enhancements exceeding 20% was inversely correlated with cloud-base distance ($r = -0.668$). The relationship remained negative but weaker in October ($r = -0.316$). August contained numerous strong-field intervals but only one TGE exceeding 20% during the analyzed period because cloud bases remained far above the station. Strong near-surface electric fields alone are therefore insufficient for frequent TGE detection when the atmospheric path between the acceleration region and the detectors is too long. No systematic decrease in cloud-base distance was found during 2012–2024; consequently, interannual changes in TGE occurrence cannot be attributed to a monotonic lowering of the cloud base.

Cloud-base distance also affects the particle composition observed at ground level. Gamma rays can propagate through considerably greater atmospheric depths than electrons, whereas electron-rich TGEs require the lower boundary of the acceleration region to approach within several tens of meters of the station. The relatively low cloud bases observed in May and October are therefore favorable for electron detection, while the higher summer cloud bases favor gamma-dominated events.

The 2012–2025 series additionally reveals a statistically significant positive tendency in annual mean air temperature and a simultaneous negative tendency in measured daytime solar radiation. The air temperature slope is $+0.116 \pm 0.029\ ^{\circ}\mathrm{C\,yr^{-1}}$, equivalent to $1.16\ ^{\circ}$Cper decade, whereas daytime solar radiation decreases by $2.62 \pm 0.79\ \mathrm{W\,m^{-2}\,yr^{-1}}$. Bootstrap confidence intervals remain entirely positive for temperature and entirely negative for solar radiation, and both signs are preserved after removing the most extreme observations. The temperature increase was not accompanied by an increase in measured incoming daytime short-wave irradiance.

The principal result of this study is the quantitative connection established between atmospheric conditions, near-surface electric fields, and high-energy particle fluxes. The Aragats measurements show that low cloud bases and strong electrification must occur together for frequent and intense TGEs to be detected, while seasonal cloud geometry strongly influences whether electrons can survive to ground level. At the same time, the air temperature and solar-radiation records document significant recent environmental tendencies at this high-altitude site. These results demonstrate the value of combining meteorological, atmospheric-electric, and particle measurements to investigate coupled processes in the mountain atmosphere.

## ACKNOWLEDGMENTS

We thank the staff of Aragats Research Station for maintaining the DAVIS weather stations, and Suren Chilingaryan for developing the ADEI data analysis platform. The authors acknowledge the support of the Science Committee of the Republic of Armenia in modernizing the technical infrastructure of high-altitude stations (Research Project No. 21AG-1C012).

**Data availability**

**The data for this paper are available in graphical and numerical formats via the multivariate** visualization software ADEI on the Cosmic Ray Division (CRD) web page of the Yerevan Physics Institute: http://adei.crd.yerphi.am/ (accessed on 1 April 2026). The ADEI infrastructure and its use for Aragats multivariate data are described in Chilingarian et al. (2024c) and ADEI (2026).

**DECLARATION OF INTEREST**

The authors declare that there are no known competing financial interests or personal relationships that could have influenced the work reported in this paper.

A.C.: Conceptualization, formal analysis, and writing—original draft; N.M. and Q.P.: Investigation and visualization.

A. Chilingarian received funding from the Science Committee of the Republic of Armenia; Grant ID 21AG-1C012.

**References**

ADEI (2026). Advanced Data Extraction Infrastructure, Cosmic Ray Division, Yerevan Physics Institute. Available at: http://adei.crd.yerphi.am/ (accessed 27 July 2026).
Boltek Corporation (2023). EFM-100C Electric Field Monitor: User Manual. Available at: https://www.boltek.com/EFM-100C_Manual_030323.pdf (accessed 27 July 2026).
Chilingarian, A., Daryan, A., Arakelyan, K. et al. 2010. Ground-based observations of thunderstorm-correlated fluxes of high-energy electrons, gamma rays, and neutrons. Physical Review D, 82, 043009. https://doi.org/10.1103/PhysRevD.82.043009

Chilingarian, A., Hovsepyan, G., & Hovhannisyan, A. (2011). Particle bursts from thunderclouds: Natural particle accelerators above our heads. Physical Review D, 83, 062001. https://doi.org/10.1103/PhysRevD.83.062001

Chilingarian, A., Hovsepyan, G., and Mailyan, B. (2017). In situ measurements of the runaway breakdown on Aragats Mountain. Nuclear Instruments and Methods in Physics Research Section A, 874, 19–27. https://doi.org/10.1016/j.nima.2017.08.022

Chilingarian, A., Hovsepyan, G., Karapetyan, T., Sargsyan, B., and Chilingaryan, S. (2022). Measurements of energy spectra of relativistic electrons and gamma-rays from avalanches developed in the thunderous atmosphere with the Aragats Solar Neutron Telescope. Journal of Instrumentation, 17, P03002. https://doi.org/10.1088/1748-0221/17/03/P03002

Chilingarian A., G. Hovsepyan (2023) Proving ‘‘new physics’’ by measuring cosmic ray fluxes, Astronomy and Computing 44, 100714. https://doi.org/10.1016/j.ascom.2023.100714

Chilingarian, A., Hovsepyan, G., Sargsyan, B., Karapetyan, T., Aslanyan, D., and Kozliner, L. (2024a). Enormous impulsive enhancement of particle fluxes observed on Aragats on May 23, 2023. Advances in Space Research, 74, 4377–4387. https://doi.org/10.1016/j.asr.2024.02.041

Chilingarian, A., Sargsyan, B., Karapetyan, T., Aslanyan, D., Chilingaryan, S., Kozliner, L. and Khanikyanc, Y., (2024b). Extreme thunderstorm ground enhancements registered on Aragats in 2023. Physical Review D, 110, 063043. https://doi.org/10.1103/PhysRevD.110.063043

Chilingarian, A., Karapetyan, T., Sargsyan, B., Khanikyanc, Y., and Chilingaryan, S. (2024c). Measurements of particle fluxes, electric fields, and lightning occurrences at the Aragats Space-Environmental Center (ASEC). Pure and Applied Geophysics, 181, 1963-1985. https://doi.org/10.1007/s00024-024-03481-5

Chilingarian, A., Williams, E., Hovsepyan, G. & Mkrtchyan, H., 2025. Why Schonland failed in his search for runaway electrons from thunderstorms. *Journal of Geophysical Research: Atmospheres*, **130**, e2024JD042350. https://doi.org/10.1029/2024JD042350.

Chilingarian, A., Sargsyan, B., Kozliner, L., & Karapetyan, T. (2026). Resolving GLE 77 of November 11, 2025: First simultaneous recovery of neutron and muon energy spectra. *Journal of Geophysical Research: Space Physics*, *131*, e2026JA035112. https://doi.org/10.1029/2026JA035112

Cosmic Ray Division (2026a). Research Stations: Aragats. Available at: https://crd.yerphi.am/Research_Stations (accessed 27 July 2026).

Cosmic Ray Division (2026b). Weather Station: Davis Wireless Vantage Pro2 Plus at Aragats. Available at: https://crd.yerphi.am/WS (accessed 27 July 2026).

Davis Instruments (2021). Vantage Pro2 Integrated Sensor Suite and Console Manuals. Davis Instruments, Hayward, California.

Efron, B., and Tibshirani, R.J. (1993). An Introduction to the Bootstrap. Chapman & Hall/CRC, New York. https://doi.org/10.1201/9780429246593

Gurevich, A.V., Milikh, G.M., Roussel-Dupre, R. (1992). Runaway electron mechanism of air breakdown and preconditioning during a thunderstorm. Phys. Lett. A 165 (5-6), 463–468.

Lawrence, M. G. (2005). The relationship between relative humidity and the dewpoint temperature in moist air: A simple conversion and applications. Bulletin of the American Meteorological Society, 86, 225–233. https://doi.org/10.1175/BAMS-86-2-225

Samanta, S., Tyagi, B., Vissa, N. K., and Sahu, R. K. (2020). A new thermodynamic index for thunderstorm detection based on cloud base height and equivalent potential temperature. Journal of Atmospheric and Solar-Terrestrial Physics, 207, 105367. https://doi.org/10.1016/j.jastp.2020.105367

Sargsyan, B., Chilingarian, A. (2026). Spectral features of ground-level enhancement and Forbush decrease in solar cycle 25, Advances in Space Research, in press. https://doi.org/10.1016/j.asr.2026.04.068

Schonland, B.F.J. and Viljoen, J.P.T. (1933), On a penetrating radiation from thunderclouds, Proc. Roy. Soc. A, Vol. 140, No. 841, 314-333.

Wilks, D.S. (2019). Statistical Methods in the Atmospheric Sciences, 4th ed. Elsevier, Amsterdam.

Williams, E., Mailyan, B., Karapetyan, G., and Mkrtchyan, H. (2023). Conditions for energetic electrons and gamma rays in thunderstorms. Journal of Geophysical Research: Atmospheres, 128, e2023JD039612. https://doi.org/10.1029/2023JD039612

Wilson, C.T.R. (1925). The acceleration of β-particles in strong electric fields such as those of thunderclouds. Proc. Cambridge Philos. Soc. 22, 534–538, http://dx.doi.org/10.1017/S0305004100003236.